\documentclass[twocolumn,showpacs,preprintnumbers,amsmath,amssymb]{revtex4}

\usepackage{graphicx}

\usepackage{dcolumn}

\usepackage{bm}

\newcommand{\ket}[1]{|#1\rangle}

\begin{document}

\title{Detection of a persistent-current qubit by resonant activation}

\author{P. Bertet}
\affiliation{Quantum Transport Group, Kavli Institute of
Nanoscience, Delft University of Technology, Lorentzweg
$1$,$2628CJ$, Delft, The Netherlands}
\author{I. Chiorescu}
\altaffiliation[Present address: ]{Department of Physics and
Astronomy, Michigan State University, East Lansing, MI-48824,
USA.} \affiliation{Quantum Transport Group, Kavli Institute of
Nanoscience, Delft University of Technology, Lorentzweg
$1$,$2628CJ$, Delft, The Netherlands}
\author{K. Semba}
\affiliation{NTT Basic Research Laboratories, NTT Corporation,
Atsugi-shi, Kanagawa 243-0198, Japan}
\author{C. J. P. M. Harmans}
\affiliation{Quantum Transport Group, Kavli Institute of
Nanoscience, Delft University of Technology, Lorentzweg
$1$,$2628CJ$, Delft, The Netherlands}
\author{J. E. Mooij}
\affiliation{Quantum Transport Group, Kavli Institute of
Nanoscience, Delft University of Technology, Lorentzweg
$1$,$2628CJ$, Delft, The Netherlands}

\begin{abstract}

We present the implementation of a new scheme to detect the
quantum state of a persistent-current qubit. It relies on the
dependency of the measuring Superconducting Quantum Interference
Device (SQUID) plasma frequency on the qubit state, which we
detect by resonant activation. With a measurement pulse of only
$5ns$, we observed Rabi oscillations with high visibility
($65\%$).

\end{abstract}


\maketitle

Superconducting circuits containing Josephson junctions are
promising candidates to be the building blocks (qubits) for future
quantum computers \cite{Makhlin02}. Coherent dynamics has been
observed with several different qubit designs. Complex sequences
of rotations on the Bloch sphere could be performed in NMR-like
experiments : Rabi, Ramsey, spin-echo, composite pulse sequences
were demonstrated \cite{Nakamura99,Vion02,Martinis02,Chiorescu03}.
In these experiments, the detection of the qubit energy state is
done after the coherent operations have been accomplished by
reading out a macroscopic detector with an output that is
correlated to the qubit energy eigenstate. By averaging the signal
delivered, one infers the qubit eigenstate occupation probability.

A hysteretic DC SQUID has been succesfully employed to measure the
state of a flux-qubit \cite{Chiorescu03} (a similar scheme was
used in \cite{Vion02} in the case of a split Cooper pair box). In
this experiment, a fast dc current pulse (DCP) sent through the
SQUID induced switching to the finite-voltage state conditional on
the qubit state. In this Rapid Communication, we propose and
demonstrate a new read-out method, faster than the previous one,
which does not require a large bandwidth line to inject current
through the SQUID and thus will allow a stronger filtering of the
bias lines. The method is based on the dependence of the SQUID
plasma frequency on the qubit's state measured by a resonant
activation microwave pulse (RAP). We compare Rabi oscillations
obtained by the two schemes in exactly the same conditions, and
show that the RAP method significantly increases the contrast.

The persistent-current qubit consists of a micron-size
superconducting loop intersected with three Josephson junctions
\cite{Mooij99}. It is threaded by a magnetic flux $\Phi_x$
generated by an external coil. When the total phase across the
three junctions $\gamma_q$ is close to $\pi$ (meaning that
$\Phi_x$ is close to $\Phi_0/2$), the loop has two low-energy
eigenstates (ground state $0$ and excited state $1$) well
separated from the higher-energy ones, and can thus be used as a
qubit. They are linear combinations of two states $\ket{ \uparrow
}$ and $\ket{ \downarrow}$ which carry an average
persistent-current of $+I_p$ and $-I_p$ and are coupled by
tunnelling. The effective hamiltonian reads $H=-h (\epsilon \sigma
_z/2 - \Delta \sigma _x /2)$ in the $\{ \ket{ \uparrow }, \ket{
\downarrow} \}$ basis, where $h \Delta /2$ is the tunnelling
matrix element between the two basis states, and $\epsilon=2 I_p
(\Phi_{x}-\Phi_0/2)/h$. This leads to an energy separation between
the two states $E_1-E_0 \equiv h f_q =  h \sqrt{\Delta ^2 +
\epsilon ^2 }$ \cite{Caspar00}.

\begin{figure}
\resizebox{.48\textwidth}{!}{\includegraphics{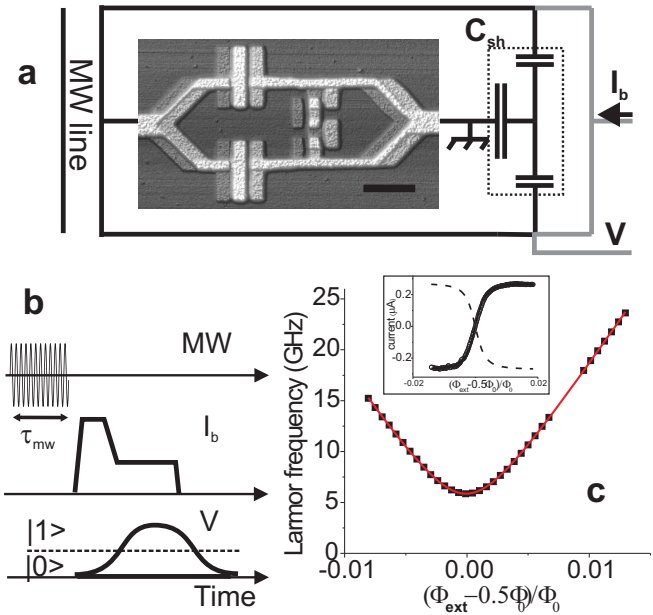}}
\caption{(a) AFM picture of the SQUID and qubit loop (the scale
bar indicates $1~\mu m$). Two layers of Aluminium were evaporated
under $\pm ~20 ~^\circ$ with an oxidation step in between. The
Josephson junctions are formed at the overlap areas between the
two images. The SQUID is shunted by a capacitor $C_{sh}=12~pF$
connected by Aluminium leads of inductance $L=170~pH$ (solid black
line). The current is injected through a resistor (grey line) of
$400 ~ \Omega$. (b) DCP measurement method : the microwave pulse
induces the designed Bloch sphere rotations. It is followed by a
current pulse of duration $20~ns$, whose amplitude $I_b$ is
optimized for the best detection efficiency. A $400~ns$
lower-current plateau follows the DCP and keeps the SQUID in the
running-state to facilitate the voltage pulse detection. (c)
Larmor frequency of the qubit and (insert) persistent-current
versus external flux. The squares and (insert) the circles are
experimental data. The solid lines are numerical adjustments
giving the tunnelling matrix element $\Delta$, the
persistent-current $I_p$ and the mutual inductance $M$.
 \label{fig1}}
\end{figure}

The average current in state $i$ ($i=0,1$) $I_i~=~-~<~\partial H /
\partial \Phi_{x}~>_i$ can be computed from the previous
equations. It is shown in figure \ref{fig1}c for the parameters of
our sample. The qubit is inductively coupled to a hysteretic
DC-SQUID whose critical current depends on $I_i$ by an amount $M
I_i \alpha$, where $M$ is the mutual inductance between the qubit
and the SQUID, and $\alpha =
\partial I_C / \partial \Phi_{x}$ is the slope of the SQUID modulation
curve. The requirement imposed upon any detection scheme is to
detect the small (about $2\%$) variation in the SQUID critical
current associated with a transition between the qubit states in a
time shorter than the qubit's energy relaxation time $T_1$. A
first method was demonstrated in \cite{Vion02},
\cite{Chiorescu03}. It consists of applying a short DCP to the
SQUID at a value $I_b$ during a time $\Delta t$ , so that the
SQUID will switch out of its zero-voltage state
(\cite{Cottet02},\cite{Balestro03}) with a probability
$P_{sw}(I_b)$. For well-chosen parameters, the detection
efficiency $D=max_{I_b} |P_{sw}^{1}-P_{sw}^{0}|$, where
$P_{sw}^{0}$ ($P_{sw}^{1}$) is the switching probability if the
qubit is in the ground (excited) state, can approach $1$. The
switching probability then directly measures the qubit's energy
level population.

The $Al/AlO_x/Al$ tunnel junctions as well as the SQUID and
qubit's aluminium loops were obtained by e-beam lithography and
shadow-evaporation on an oxidized silicon substrate. The SQUID was
shunted by an on-chip aluminium capacitor and connected to the
current injection line through a $400~\Omega$ on-chip gold
resistor (see figure \ref{fig1}a). Both current and voltage line
were filtered by copper-powder filters \cite{Devoret87}. The
voltage across the SQUID was amplified at room-temperature and
sampled by an acquisition card. By repeating the measurements
typically $10000$ times at a rate of $25~kHz$ we derive the
switching probability. Pulsed microwave irradiation was applied to
the qubit through a coaxial cable attenuated by $20~dB$ at
$T=1.5~K$. The sample was enclosed in a copper box thermally
anchored to the mixing chamber of our dilution refrigerator (base
temperature $25~mK$) to shield it from RF noise.

\begin{figure}
\resizebox{.48\textwidth}{!}{\includegraphics{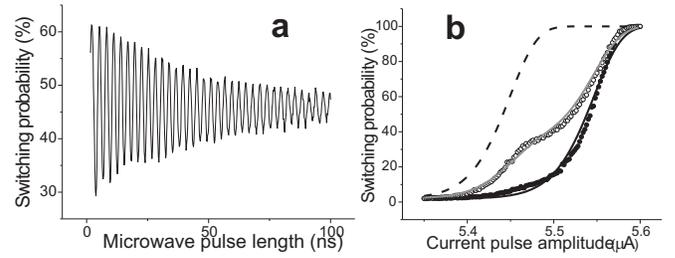}}
\caption{(a) Rabi oscillations at a Larmor frequency
$f_q=7.15~GHz$ (b) Switching probability as a function of current
pulse amplitude $I$ without (closed circles, curve $P_{sw}^0 (I)$)
and with (open circles, curve $P_{sw}^{\pi} (I_b)$) a $\pi$ pulse
applied. The solid black line $P_{th}^0 (I_b)$ is a numerical
adjustment to $P_{sw}^0 (I_b)$ assuming escape in the thermal
regime. The dotted line (curve $P_{th}^1 (I_b)$) is calculated
with the same parameters for a critical current $100nA$ smaller,
which would be the case if state $1$ was occupied with probability
unity. The grey solid line is the sum $0.32 P_{th}^1 (I_b) + 0.68
P_{th}^0 (I_b)$.
 \label{fig2}}
\end{figure}

The parameters of our qubit were determined by fitting
spectroscopic measurements with the above formulae. For
$\Delta=5.855~GHz$, $I_p=272~nA$, the agreement is excellent (see
figure \ref{fig1}c). We also determined the coupling constant
between the SQUID and the qubit by fitting the qubit ``step"
appearing in the SQUID's modulation curve \cite{Caspar00} (see
insert of figure \ref{fig1}c) and found $M=20~pH$. We first
performed Rabi oscillation experiments with the DCP detection
method (figure \ref{fig1}b). We chose a bias point $\Phi_{x}$,
tuned the microwave frequency to the qubit resonance and measured
the switching probability as a function of the microwave pulse
duration $\tau_{mw}$. The observed oscillatory behavior (figure
\ref{fig2}a) is a proof of the coherent dynamics of the qubit. A
more detailed analysis of its damping time and period will be
presented elsewhere ; here we focus on the amplitude of these
oscillations.

In the data shown here, the contrast is $32\%$, but we observed a
strong dependence on the exact bias point, ranging between $5\%$
and $40\%$. This visibility was lower than in our previous
experiment \cite{Chiorescu03}. We checked that the amplitude $I_b$
of the current pulse was optimal by measuring the switching
probability as a function of $I_b$ both without microwave
($P_{sw}^0 (I_b)$) and after a $\pi$ pulse ($P_{sw}^{\pi} (I_b)$).
Typical data are shown in figure \ref{fig2}b. The solid black
curve $P_{th}^0 (I_b)$ is a fit to $P_{sw}^0 (I_b)$ assuming that
switching occurs by thermal activation across the potential
barrier \cite{Balestro03}. The dotted line $P_{tw}^\pi (I_b)$ was
calculated with the same parameters assuming that the qubit is
always in $1$. The curve $P_{sw}^{\pi} (I_b)$ is then well
approximated by a weighted sum of the curves $P_{th}^\pi (I)$ and
$P_{th}^0 (I_b)$ with coefficients $p_g=0.68$ and $p_e=0.32$ (grey
solid line). From this we conclude that the detection efficiency
$D$ of our detector, as designed, should be of at least $85\%$,
but that the excited state of our qubit is only populated up to
$32\%$ when switching to the running state occurs at the end of
the DCP. Thus, it seems that the qubit partially relaxes towards
its ground state during the measurement process. As we will show
below, this problem does not occur with the alternative
measurement scheme described hereafter.

The SQUID, connected to the capacitor $C_{sh}$ by aluminum wires
of inductance $L$, behaves as an oscillator with a characteristic
frequency called the plasma frequency $2 \pi
f_p=[(L+L_J)C_{sh}]^{-1/2}$. This frequency depends on the bias
current $I_b$ and on the critical current $I_C$ via the Josephson
inductance $L_J=\Phi_0 / (2 \pi I_C \sqrt{1-I_b^2/I_C^2})$. Thus,
the plasma frequency takes different values $f_p^{(0)}$ or
$f_p^{(1)}$ depending on the qubit's state. This effect has
already been observed in \cite{Lupascu03} by measuring the
transmission of a weak probe at a frequency close to $f_p$ while
keeping the SQUID in the zero-voltage state. With a probe of
larger power, it has recently been demonstrated that a Josephson
junction could switch between two distinct oscillation states
while staying in the zero-voltage regime, due to its non-linearity
\cite{Devoret03bis}. Such a bifurcation amplifier could be used to
detect the state of a charge qubit \cite{Devoret03}. Here we apply
a microwave pulse at a frequency close to $f_p^{0}$. We adjust the
power so that the SQUID switches to the finite voltage state by
resonant activation \cite{Devoret84} if the qubit is in state
$\ket{0}$, whereas it stays in the zero-voltage state if it is in
state $\ket{1}$. If the two resonant activation peaks
corresponding to the frequencies $f_p^{0}$ and $f_p^{1}$ are
distinct enough, the switching probability of the SQUID measures
the probability that the qubit is in the state $0$. Similarly, in
\cite{Martinis02}, the quantum state of a current-biased Josephson
junction (phase-qubit) has been detected by applying a microwave
pulse that induces switching determined by the qubit state. In our
experiment, the detector is distinct from the qubit. Although
switching to the finite-voltage state is a dissipative process
potentially harmful for quantum computation, the specific interest
of this scheme is that it allows to heavily filter the
bias-current line since no fast DC or RF pulse is needed there.

\begin{figure}
\resizebox{.48\textwidth}{!}{\includegraphics{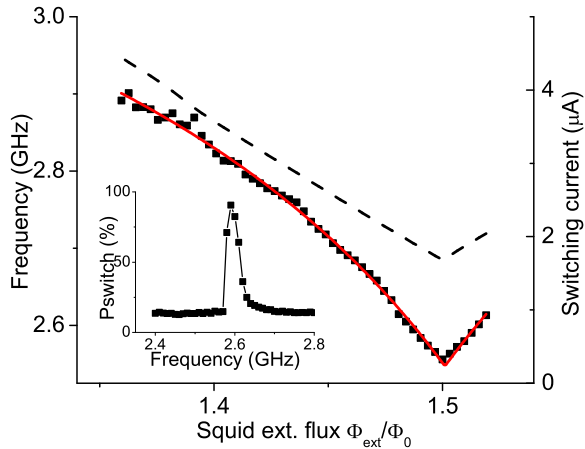}}
\caption{(insert) Typical resonant activation peak (width
$40~MHz$), measured after a $50~ns$ microwave pulse. Due to the
SQUID non-linearity, it is much sharper at low than at high
frequencies. (figure) Center frequency of the resonant activation
peak as a function of the external magnetic flux (squares). It
follows the switching current modulation (dashed line). The solid
line is a fit yielding the values of the shunt capacitor and stray
inductance given in the text.\label{fig3}}
\end{figure}

A typical resonant activation peak is shown in the insert of
figure \ref{fig3}. Its width depends on the frequency, ranging
between $20$ and $50~MHz$. This corresponds to a quality factor
between $50$ and $150$. The peak has an asymmetric shape, with a
very sharp slope on its low-frequency side and a smooth
high-frequency tail, due to the SQUID non-linearity. We could
qualitatively recover these features by simple numerical
simulations using the RCSJ model \cite{Devoret87b}. The resonant
activation peak can be unambiguously distinguished from
environmental resonances by its dependence on the magnetic flux
threading the SQUID loop $\Phi_{sq}$. Figure \ref{fig3} shows the
measured peak frequency for different fluxes around
$\Phi_{sq}=1.5\Phi_0$, together with the measured switching
current (dashed line). The solid line is a numerical fit to the
data using the above formulae. From this fit we deduce the
following values $C_{sh}=12 \pm 2 pF$ and $L=170 \pm 20 pH$, close
to the design. We are thus confident that the observed resonance
is due to the plasma frequency.

\begin{figure}
\resizebox{.48\textwidth}{!}{\includegraphics{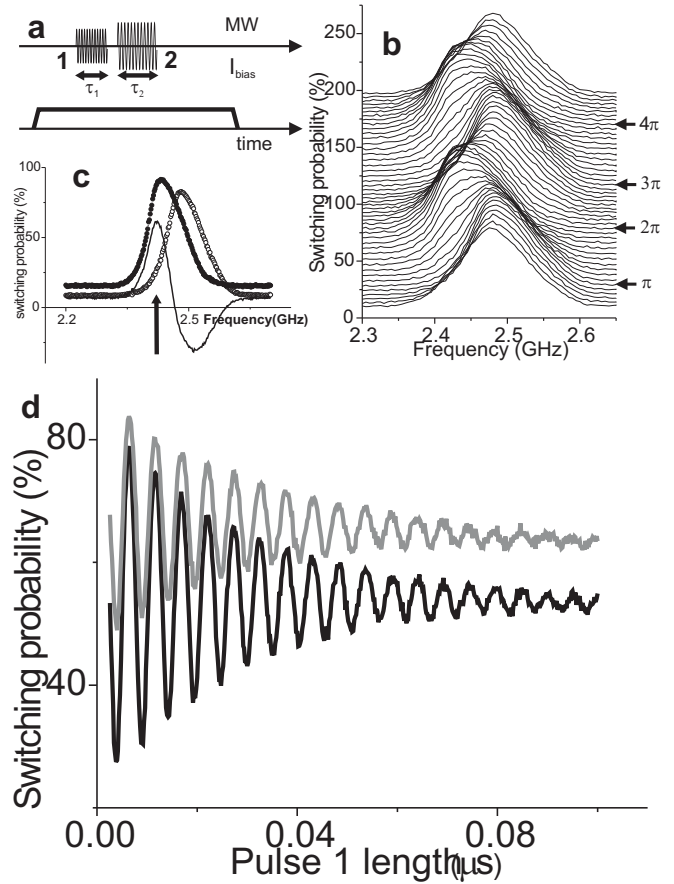}}
\caption{(a) Principle of the detection scheme. After the Rabi
pulse, a microwave pulse at the plasma frequency resonantly
enhances the escape rate. The bias current is maintained for
$500ns$ above the retrapping value. (b) Resonant activation peak
for different Rabi angle. Each curve was offset by $5\%$ for
lisibility. The Larmor frequency was $f_q=8.5~GHz$. Pulse $2$
duration was $10~ns$. (c) Resonant activation peak without (full
circles) and after (open circles) a $\pi$ pulse. The continuous
line is the difference between the two switching probabilities.
(d) Rabi oscillation measured by DC current pulse (grey line,
amplitude $A=40\%$) and by resonant activation method with a
$5~ns$ RAP (black line, $A=62\%$), at the same Larmor
frequency.\label{fig4}}
\end{figure}

We then measure the effect of the qubit on the resonant activation
peak. The principle of the experiment is sketched in figure
\ref{fig4}a. A first microwave pulse at the Larmor frequency
induces a Rabi rotation by an angle $\theta_1$. A second microwave
pulse of duration $\tau_2=10ns$ is applied immediately after, at a
frequency $f_2$ close to the plasma frequency, with a power high
enough to observe resonant activation. In this experiment, we
apply a constant bias current $I_b$ through the SQUID ($I_b=2.85
\mu A$, $I_b/I_C=0.85$) and maintain it at this value $500~ns$
after the microwave pulse to keep the SQUID in the running state
for a while after switching occurs. This allows sufficient voltage
to build up across the SQUID and makes detection easier, similarly
to the plateau used at the end of the DCP in the previously shown
method. At the end of the experimental sequence, the bias current
is reduced to zero in order to retrap the SQUID in the
zero-voltage state. We measured the switching probability as a
function of $f_2$ for different Rabi angles $\theta_1$. The
results are shown in figure \ref{fig4}b. After the microwave
pulse, the qubit is in a superposition of the states $0$ and $1$
with weights $p_0=cos^2(\theta_1/2)$ and $p_1=sin^2(\theta_1/2)$.
Correspondingly, the resonant activation signal is a sum of two
peaks centered at $f_p^{0}$ and $f_p^{1}$ with weights $p_0$ and
$p_1$, which reveal the Rabi oscillations.

We show the two peaks corresponding to $\theta_1=0$ (curve
$P_{sw}^{0}$, full circles) and $\theta_1=\pi$ (curve
$P_{sw}^{\pi}$, open circles) in figure \ref{fig4}c. They are
separated by $f_p^{0}-f_p^{1}=50~MHz$ and have a similar width of
$90~MHz$. This is an indication that the $\pi$ pulse efficiently
populates the excited state (any significant probability for the
qubit to be in $0$ would result into broadening of the curve
$P_{sw}^{\pi}$), and is in strong contrast with the results
obtained with the DCP method (figure \ref{fig2}b). The difference
between the two curves $S(f)=P_{sw}^{0}-P_{sw}^{\pi}$ (solid line
in figure \ref{fig4}c) gives a lower bound of the excited state
population after a $\pi$ pulse. Because of the above mentioned
asymmetric shape of the resonant activation peaks, it yields
larger absolute values on the low- than on the high-frequency side
of the peak. Thus the plasma oscillator non-linearity increases
the sensitivity of our measurement, which is reminiscent of the
ideas exposed in \cite{Devoret03}. On the data shown here, $S(f)$
attains a maximum $S_{max}=60\%$ for a frequency $f_2^*$ indicated
by an arrow in figure \ref{fig4}c. The value of $S_{max}$ strongly
depends on the microwave pulse duration and power. The optimal
settings are the result of a compromise between two constraints :
a long microwave pulse provides a better resonant activation peak
separation, but on the other hand the pulse should be much shorter
than the qubit damping time $T_1$, to prevent loss of excited
state population. Under optimized conditions, we were able to
reach $S_{max}=68\%$.

Finally, we fixed the frequency $f_2$ at the value $f_2^*$ and
measured Rabi oscillations (black curve in figure \ref{fig4}d). We
compared this curve to the one obtained with the DCP method in
exactly the same conditions (grey curve). The contrast is
significantly improved, while the dephasing time is evidently the
same. This enhancement is partly explained by the rapid $5~ns$ RAP
(for the data shown in figure \ref{fig4}d) compared to the $30~ns$
DCP. But we can not exclude that the DCP intrinsically increases
the relaxation rate during its risetime. Such a process would be
in agreement with the fact that for these bias conditions, $T1
\simeq 100~ns$, three times longer than the DCP duration.

In conclusion, we demonstrated a new method for detecting the
state of a persistent-current qubit. We exploited the dependence
of the SQUID's plasma frequency on the qubit's state which we
measured by applying a microwave pulse that induces resonant
activation. It provided large detection efficiency even for short
microwave pulses ($5~ns$). The contrast of the Rabi oscillations
was mainly limited by the resonant activation peaks separation. By
improving the quality factor of the plasma oscillator and using
composite pulses, we might thus reach the single-shot regime while
keeping the bias current line heavily filtered and having a large
dephasing time.

We thank Y. Nakamura for direct input in this work, F. Balestro,
A. Lupascu for fruitful discussions, R. Schouten and B. ven den
Enden for technical support. P. B. acknowledges financial support
from a European Community Marie Curie fellowship.

\end{document}